\pdfoutput=1  
\def\nk{n_\mathrm{K}}
\def\acap{\\ \nonumber \\}

\def\rfr#1{Equation\,(\ref{#1})}
\def\rfrs#1#2{Equations\,(\ref{#1})--(\ref{#2})}
\def\Rfr#1{Equation\,(\ref{#1})}
\def\Rfrs#1#2{Equations\,(\ref{#1})--(\ref{#2})}
\def\derp#1#2{\rp{\partial{#1}}{\partial{#2}}}
\def\dert#1#2{\frac{{{\textrm{d}}}{#1}}{{{\textrm{d}}}{#2}}}
\def\virg#1{``#1"}
\def\eqi{\begin{equation}}
\def\eqf{\end{equation}}
\def\rp#1#2{\frac{#1}{#2}}
\def\lb#1{\label{#1}}

\def\ton#1{\left(#1\right)}
\def\qua#1{\left[#1\right]}
\def\grf#1{\left\{#1\right\}}

\RequirePackage[2020-02-02]{latexrelease}
\documentclass{aastex631}
\usepackage{morefloats}
\usepackage[title]{appendix}
\usepackage{textcomp}
\usepackage{booktabs}
\usepackage{multirow}
\usepackage{rotating,tabularx}
\usepackage{float}
\usepackage{enumerate}
\usepackage{rotating}
\usepackage[polutonikogreek,english]{babel}
\usepackage{amsmath,starfont,textgreek,w-greek}
\usepackage[flushleft]{threeparttable}
\usepackage{amsthm}
\usepackage{bigints}
\usepackage{amscd}
\usepackage[mathlines]{lineno}
\usepackage{amssymb,dsfont}
\usepackage{graphicx,epsfig}
\usepackage{txfonts}
\usepackage{xr-hyper}
\usepackage{hyperref}
\usepackage[utf8]{inputenc}
\usepackage{newunicodechar,graphicx}

\usepackage[nointegrals]{wasysym}
\usepackage[caption=false]{subfig}

\DeclareRobustCommand{\okina}{%
  \raisebox{\dimexpr\fontcharht\font`A-\height}{%
    \scalebox{0.8}{`}%
  }%
}
\newunicodechar{ʻ}{\okina}

\allowdisplaybreaks

\makeatletter
 \DeclareRobustCommand\ref{%
    \@ifstar\@refstar\T@ref
  }%
  \DeclareRobustCommand\pageref{%
    \@ifstar\@pagerefstar\T@pageref
  }%
 \makeatother

\begin{document}

\title{Post--Keplerian perturbations of the hyperbolic motion  in the field of a rotating massive object. Analysis in terms of osculating and nonosculating (contact) elements}

\shortauthors{L. Iorio}

\author[0000-0003-4949-2694]{Lorenzo Iorio}
\affiliation{Ministero dell' Istruzione e del Merito. Viale Unit\`{a} di Italia 68, I-70125, Bari (BA),
Italy}

\email{lorenzo.iorio@libero.it}

\begin{abstract}
The perturbations of the  hyperbolic motion of a test particle due to the general relativistic gravitoelectromagnetic Schwarzschild and Lense--Thirring  components of the gravitational field of a  rotating massive body  are analytically worked out to the first post--Newtonian level in terms of the osculating Keplerian orbital elements. To the Newtonian order, the impact of the quadrupole mass moment of the source is calculated as well. The resulting analytical expressions are valid for a generic orientation in space of both the orbital plane of the probe and the spin axis of the primary, and for arbitrary values of the eccentricity. They are applied  to ʻOumuamua, an interstellar asteroid which recently visited our solar system along an unbound heliocentric orbit, and to the Near Earth Asteroid Rendezvous (NEAR) spacecraft during its flyby of the Earth.
The calculational approach  developed can be straightforwardly extended to any alternative models of gravity as well.
\end{abstract}


\section{Introduction}
Let  a localized gravitational source like, e.g., a planet, a natural satellite, a main sequence star or any astrophysical compact object endowed with mass $M$, equatorial radius $R_\mathrm{e}$, quadrupole mass moment $J_2$ and angular momentum $\boldsymbol{J}$ be considered.  Let its external gravitational field be calculated in points far enough so that it is weak and the speeds of any moving test particles are small with respect to the speed of light in vacuum $c$. Then, in addition to the dominant Newtonian inverse--square mass monopole,  also further post--Keplerian (pK) terms of both Newtonian and post--Newtonian (pN) origin come into play. The most relevant ones are the classical contribution of $J_2$  and, to the first post--newtonian (1pN) order, the so--called gravitoelectromagnetic  Schwarzschild and Lense--Thirring (LT) components induced by $M$ and $\boldsymbol{J}$, respectively.

Until now, their orbital effects have been studied mainly in the case of bound, otherwise Keplerian elliptical trajectories \citep{1991ercm.book.....B,Sof89,2011rcms.book.....K,2016ASSL..436.....G,SoffelHan19,OLeary21,2024gpno.book.....I}, used as tools to perform tests of gravitational theories. The most famous case is represented by the then anomalous perihelion precession of Mercury of $42.98$ arcseconds per century (arcsec cty$^{-1}$) \citep{1986Natur.320...39N} in the field of the Sun, known since the second half of the nineteenth century \citep{LeVer1859,LeV2}, and its successive  explanation by Einstein \citep{Ein15} in terms of his newborn General Theory of Relativity (GTR). For a historical overview, see, e.g., \citet{1982mpfl.book.....R}.

Instead, studies of pK perturbations of hyperbolic trajectories are comparatively much more rare, being mainly focussed on the effects of the primary's oblateness  for a particular orientation\footnote{Indeed, since they are generally devoted to flybys of Earth, whose spin axis is well known,  the reference $z$ axis is aligned just with the latter.} of its spin axis $\boldsymbol{\hat{J}}$ \citep{Sauer63,1999Icar..138..309A,2001Icar..150..168R,MartiGurf13,2015JGCD...38.1690K}. Other works investigated the hyperbolic motions of test particles and photons in the Schwarzschild spacetime at various levels of completeness \citep{Morton21,1930JaJAG...8...67H,Leavitt39,1959RSPSA.249..180D,1961RSPSA.263...39D,Mieln62,1980AuJPh..33..757D,2010arXiv1008.1964H,2024PhRvD.109l4056C}. The case of the hyperbolic motion of a spinning particle in the Schwarzschild metric was treated by \citet{2017GReGr..49...84B}, while \citet{2022EPJC...82.1088B} dealt with geodesic motion in Euclidean Schwarzschild geometry. To the author's knowledge, the gravitomagnetic effects of the rotation of the primary on hyperbolic trajectories have never been treated so far, apart from the study by \citet{2023PhRvD.107l4058M} in the Kerr metric.

Flybys of planets and natural satellites by artificial spacecraft traveling along patched hyperbolic conical sections are commonplace in current astrodynamics and planetary sciences \citep{Flandro66,And97,2003AmJPh..71..448V,2007NewA...12..383A}. Furthermore, they are often repeated several times within the same missions; suffice it to think about the grand tour of the Cassini probe in the Kronian system \citep{Wolf95}.
Finally, also the Galactic Centre and the cluster of stars surrounding the supermassive black hole at Sgr A$^\ast$ \citep{2010RvMP...82.3121G} may be considered. Indeed, the star S111 is following a hyperbolic path \citep{2008A&A...492..419T,2009ApJ...692.1075G,2017ApJ...837...30G}; other stars like that might be discovered in the future.
Eventually, such kind of trajectories may represent, in principle,  further opportunities to test gravitational theories in addition to the traditional bound ones.

Here, in order to make closer contact with observations in  actually accessible astronomical scenarios, a perturbative approach is followed. It allows to analytically calculate the variations experienced by all the usual  Keplerian orbital elements of a hyperbolic trajectory perturbed by the aforementioned pK components of the gravitational field of the primary. In this respect, the present work follows a similar strategy as that adopted in \citet{Sauer63,1999Icar..138..309A,2001Icar..150..168R,2015JGCD...38.1690K}. Nonetheless, the effects of the Newtonian quadrupole mass moment $J_2$ and of the 1pN gravitomagnetic LT field are worked out in full generality, without any a priori simplifying assumptions about the  orientations of both $\boldsymbol{\hat{J}}$ and the orbit in space. Furthermore, all the formulas obtained are valid for any values of the eccentricity.

The paper is organized as follows.
In Section~\ref{sec:calc}, the basics of the Keplerian hyperbolic motion is reviewed and the perturbative equations for the rates of change of the Keplerian orbital elements in the form of Lagrange are presented for such kind of unperturbed, reference trajectory. Furthermore, the way of calculating the disturbing function, to be used with the aforementioned equations, for the pK effects considered  is discussed as well. The 1pN gravitoelectric shifts induced solely by $M$ are calculated in Section~\ref{sec:GE}. The 1pN gravitomagnetic LT perturbations due to $\boldsymbol{J}$ is the subject of
Section~\ref{sec:LT}, while the impact of $J_2$ is worked out, to the Newtonian order, in Section~\ref{sec:J2}; both effects are calculated without any a priori assumptions on both $\boldsymbol{\hat{J}}$ and the orientation of the orbital plane. \textcolor{black}{Certain subtleties concerning the proper use of the Lagrange planetary equations in presence of velocity--dependent disturbing functions are dealt with in Section~\ref{sec:mike}.} The results of the previous Sections are used for numerical calculation in Section~\ref{sec:num} for two astronomical scenarios in our solar system: the interstellar asteroid ʻOumuamua and the Sun in Section~\ref{sec:Oum}, and the spacecraft Near Earth Asteroid Rendezvous (NEAR) approaching the Earth in Section~\ref{sec:Near}. Section~\ref{Sec:end} summarizes the findings  and offers conclusions.
\section{Calculational overview}\lb{sec:calc}
In the Keplerian hyperbolic motion \citep{2001Icar..150..168R,2005ormo.book.....R,2016ASSL..436.....G}, $a$ is the semimajor axis, $e$ is the eccentricity, $I$ is the inclination, $\Omega$ is the longitude of the ascending node, $\omega$ is the argument of pericentre, and $\eta$ is the mean anomaly at epoch. The semimajor axis measures the distance between the vertex, namely the point Q of closest approach to the primary, and the centre O of the hyperbola; it is $a<0$. For the eccentricity, which is at any time the constant ratio of the distance of the test particle at the point P$\ton{t}$ on the hyperbola to the focus F where the primary resides to the distance of P$\ton{t}$ itself to the directrix, it always holds $e>1$; the larger it is, the straightest the hyperbola, while its asymptotes tend to get closer for $e\gtrsim 1$. The inclination is the tilt of the orbital plane to the reference $\grf{x,\,y}$ plane of the body--fixed reference frame adopted. The longitude of the ascending node is the angle, counted in the reference plane, from the reference $x$ direction to the point N on the line of nodes crossed by the test particle from below; the line of nodes is the intersection between the orbital and the fundamental planes. The argument of pericentre is the angle, reckoned in the orbital plane, from N to Q. The mean anomaly at epoch is proportional to the time of closest approach $t_\mathrm{p}$; indeed, from the definition of the mean anomaly
\eqi
\mathcal{M}\ton{t} = \nk\ton{t - t_\mathrm{p}} = \nk t + \eta\lb{Manom},
\eqf
it follows
\eqi
\eta:= -\nk t_\mathrm{p}\lb{tperi}.
\eqf
In \rfrs{Manom}{tperi},
\eqi
\nk = \sqrt{-\rp{\upmu}{a^3}}
\eqf
is the Keplerian mean motion which, of course, has not the same meaning as for the elliptic orbits. Moreover,
\eqi
\upmu:=GM
\eqf
is the standard gravitational parameter of the source of the gravitational field given by the product of its mass by the Newtonian gravitational constant $G$. Instead, $I,\,\Omega$ and $\omega$ determine the orientation of the orbit in space and of the orbit itself within its orbital plane also for the hyperbolic motion.

In view of the forthcoming calculation, it is convenient to express the mean anomaly in terms of the hyperbolic eccentric anomaly $H\ton{t}$ as \citep{2001Icar..150..168R}
\eqi
\mathcal{M} \lb{ManomH} = e\sinh H - H.
\eqf
Furthermore, it is \citep{2001Icar..150..168R}
\begin{align}
\sinh H \lb{sinhH} &= \rp{\sin f\sqrt{e^2 - 1}}{1 + e\cos f}, \acap
\cosh H \lb{coshH} & = \rp{e + \cos f}{1 + e\cos f}.
\end{align}
In \rfrs{sinhH}{coshH}, $f\ton{t}$ is the true anomaly, counted from Q to P$\ton{t}$ in such a way that $f=0$ at the pericentre and
\eqi
-f_\infty \leq f \leq f_\infty,\lb{Df}
\eqf
where
\eqi
f_\infty = \arccos\ton{-\rp{1}{e}}.\lb{finfty}
\eqf
From \rfr{Manom} and by using \rfrs{ManomH}{coshH}, one gets
\eqi
\dert{t}{f} = \rp{\ton{e^2 -1}^{3/2}}{\nk\ton{1 + e\cos f}^2}.\lb{dtdf}
\eqf
The instantaneous distance of the test particle from the primary can be expressed as \citep{2001Icar..150..168R}
\eqi
r = a\ton{1 - e\cos H}.\lb{rhyp}
\eqf
The position and velocity vectors, referred to the orbital plane\footnote{In the orbit--fixed frame, the $X$ axis is directed along the line of apsides towards the pericentre.} $\grf{X,\,Y}$, are \citep{2001Icar..150..168R}
\begin{align}
\boldsymbol{r} \lb{poshyp}& = \grf{a\ton{\cos H - e},\,\sqrt{-a p}\sin H,\,0}, \acap
\boldsymbol{v} \lb{velhyp}& = \grf{\rp{\sqrt{-\upmu a}}{r}\sin H,\,\rp{\sqrt{\upmu p}}{r}\cos H,\,0},
\end{align}
where
\eqi
p := -a\ton{e^2 - 1}\lb{slrhyp}
\eqf
is the semilatus rectum.
The hyperbolic excess velocity is defined as \citep{2001Icar..150..168R}
\eqi
v_\infty = -\nk a=\sqrt{-\rp{\upmu}{a}}.\lb{vinft}
\eqf

\Rfrs{sinhH}{coshH} allow to express \rfrs{rhyp}{velhyp} in terms of $f$.

The components of $\boldsymbol{r}$ and $\boldsymbol{v}$ can be referred to the primary--fixed reference frame by means of the rotation matrix \citep{2000Monte}
\eqi
\mathcal{R}\ton{\Omega,I,\omega}=\mathcal{R}_z\ton{-\Omega}\mathcal{R}_x\ton{-I}\mathcal{R}_z\ton{-\omega},\lb{Rotam}
\eqf
where, for a generic angle $\phi$, it is
\begin{align}
\mathcal{R}\ton{-\phi}_z &=\left(
  \begin{array}{ccc}
    \cos\phi & -\sin\phi & 0 \\
   \sin\phi  & \cos\phi & 0 \\
    0 & 0 & 1 \\
  \end{array}
\right),\acap
\mathcal{R}\ton{-\phi}_x &=\left(
  \begin{array}{ccc}
    1 & 0 & 0\\
    0 & \cos\phi  & -\sin\phi\\
    0 & \sin\phi  & \cos\phi\\
  \end{array}
\right).
\end{align}
\Rfr{Rotam} allows to determine the orientation of the orbit in space and of the orbit itself within the orbital plane in full generality.

For calculational purposes, it is convenient to introduce the following mutually orthogonal unit vectors \citep{Sof89,1991ercm.book.....B,SoffelHan19}
\begin{align}
\boldsymbol{\hat{l}}\lb{elle}&:=\grf{\cos\Omega,\,\sin\Omega,\,0},\acap
\boldsymbol{\hat{m}}\lb{emme}&:=\grf{-\cos I\sin\Omega,\,\cos I\cos\Omega,\,\sin I},\acap
\boldsymbol{\hat{h}}\lb{acca}&:=\grf{\sin I\sin\Omega,\,-\sin I\cos\Omega,\,\cos I};\acap
\end{align}
$\boldsymbol{\hat{l}}$ is directed along the line of nodes towards the ascending node, $\boldsymbol{\hat{h}}$  is perpendicular to the orbital plane, being aligned with the orbital angular momentum, and $\boldsymbol{\hat{m}}$ lies in the orbital plane so that
\eqi
\boldsymbol{\hat{l}} \boldsymbol{\times}\boldsymbol{\hat{m}} = \boldsymbol{\hat{h}}
\eqf
holds.

The planetary equations in the form of Lagrange which allow to calculate the perturbations of the Keplerian orbital elements in the case of  the hyperbolic motion are
\citep{2001Icar..150..168R}
\begin{align}
\dert a t \lb{dadt}& = -\rp{2}{\nk a}\derp{\mathfrak{R}}{\eta},\acap
\dert e t & = \rp{\sqrt{e^2 - 1}}{\nk a^2 e}\derp{\mathfrak{R}}{\omega} + \rp{\ton{e^2 - 1}}{\nk a^2 e}\derp{\mathfrak{R}}{\eta},\acap
\dert I t & =  -\rp{1}{\nk a^2\sqrt{e^2 - 1}\sin I}\derp{\mathfrak{R}}{\Omega} + \rp{\cos I}{\nk a^2\sqrt{e^2 - 1}\sin I}\derp{\mathfrak{R}}{\omega},\acap
\dert \Omega t & =  \rp{1}{\nk a^2\sqrt{e^2 - 1}\sin I}\derp{\mathfrak{R}}{I},\acap
\dert \omega t & =  -\rp{\sqrt{e^2 - 1}}{\nk a^2 e}\derp{\mathfrak{R}}{e} - \rp{\cos I}{\nk a^2\sqrt{e^2 - 1}\sin I}\derp{\mathfrak{R}}{I},\acap
\textcolor{black}{\dert \eta t} \lb{detadt}& =  \textcolor{black}{\rp{2}{\nk a}\derp{\mathfrak{R}}{a} - \rp{\ton{e^2 - 1}}{\nk a^2 e}\derp{\mathfrak{R}}{e}},
\end{align}
where  $\mathfrak{R}$ is the disturbing function.
$\mathfrak{R}$ is given by the pK \textcolor{black}{part} \textcolor{black}{$\mathcal{L}^\mathrm{pK}$} of the Lagrangian per unit mass $\mathcal{L}$ of the test particle which can be obtained from the spacetime metric tensor $g_{\mu\nu},\,\mu,\nu=0,1,2,3$ as follows.

Written in spatially isotropic or harmonic coordinates, the latter can be expressed, to the pN order, as
\begin{align}
g_{00} \lb{g00} & \simeq 1 + h_{00} = 1 + \rp{2U\ton{\boldsymbol{r}}}{c^2} + \rp{2 U^2\ton{\boldsymbol{r}}}{c^4} + \mathcal{O}\ton{1/c^6}, \acap
g_{0i} \lb{g0i} & \simeq h_{0i} = \mathcal{O}\ton{1/c^3},\,i=1,2,3, \acap
g_{ij} \lb{gij} & \simeq -1 + h_{ij} = -\qua{1 - \rp{2U\ton{\boldsymbol{r}}}{c^2}}\delta_{ij} + \mathcal{O}\ton{1/c^4},\,i,j=1,2,3.
\end{align}
In \rfrs{g00}{gij}, the coefficients $h_{\mu\nu},\,\mu,\nu=0,1,2,3$ are the pN corrections to the constant components $\eta_{\mu\nu},\,\mu,\nu=0,1,2,3$ of the \virg{flat} Minkowskian spacetime metric tensor,
\eqi
\delta_{ij} := \left\{
\begin{aligned}
& 1\,\mathrm{for}\,i=j \\
& 0\,\mathrm{for}\,i\neq j,
\end{aligned}
\right.\,i,j=1,2,3,
\eqf
is the 3--dimensional Kronecker delta  \citep{nist010}, and $U\ton{\boldsymbol{r}}$ is the Newtonian potential of the source including also $J_2$
\eqi
U\ton{\boldsymbol{r}} = -\rp{\upmu}{r}\qua{1 - J_2\ton{\rp{R_\mathrm{e}}{r}}^2\mathcal{P}_2\ton{\boldsymbol{\hat{J}}\boldsymbol\cdot\boldsymbol{\hat{r}}}},\lb{Upot}
\eqf
in which
\eqi
\mathcal{P}_2\ton{\xi} = \rp{3\xi^2 - 1}{2}
\eqf
is the Legendre polynomial of degree $\ell =2 $ in the generic dimensionless  argument $\xi$.
Furthermore,
\eqi
h_{0i}  = \rp{2GJ\epsilon_{ijk}\hat{J}^j x^k}{c^3r^3}, \,i=1,2,3\lb{h0i}
\eqf
where
\eqi
\epsilon_{ijk}:==\left\{
              \begin{array}{lll}
                +1 & \hbox{if $\ton{i,\,j,\,k}$ is $\ton{1,\,2,\,3}$,\,$\ton{2,\,3,\,1}$, or~$\ton{3,\,1,\,2}$  } \\
                -1 & \hbox{if $\ton{i,\,j,\,k}$ is $\ton{3,\,2,\,1}$,\,$\ton{1,\,3,\,2}$, or~$\ton{2,\,1,\,3}$ } \\
                0 & \hbox{if $i=j$, or~$j=k$, or~$k=i$}
              \end{array}
            \right.
\eqf
is the 3--dimensional Levi--Civita symbol \citep{nist010},
are the components of the gravitomagnetic LT potential. In \rfr{h0i}, $\hat{J}^i,\,i=1,2,3$ are the components of the spin unit vector $\boldsymbol{\hat{J}}$, and $x^k,\,k=1,2,3$ are the Cartesian coordinates $x,\,y,\,z$ of the test particle. In \rfr{h0i}, the Einstein summation convention \citep{nist010} is applied to the dummy summation indexes $j$ and $k$.

To the 1pN order, the Lagrangian per unit mass turns out to be
\citep[p.\,56,\,Equation\,(2.2.53)]{1991ercm.book.....B}
\eqi
\mathcal{L} = \mathcal{L}_\mathrm{N} + \mathcal{L}^\mathrm{1pN},
\eqf
where\footnote{Here, the velocity  components $v^i,\,i=1,2,3$ are calculated with respect to the coordinate time $t$ \citep{1991ercm.book.....B}.}
\begin{align}
\mathcal{L}_\mathrm{N} \lb{LNt}&= \rp{1}{2}v^2 - \rp{1}{2}c^2 h^{\ton{1/c^2}}_{00}, \acap
\mathcal{L}^\mathrm{1pN} \lb{LpN} & = - \rp{1}{2}c^2 h^{\ton{1/c^4}}_{00} + \rp{v^4}{8c^2} - \rp{1}{4}h_{00}v^2 + \rp{c^2}{8}h^2_{00} - \rp{1}{2}h_{ij}v^i v^j - ch_{0j} v^j,
\end{align}
where $h_{\alpha\beta},\,\alpha,\beta=0,1,2,3$ are given by \rfrs{g00}{gij}; $h_{00}^{\ton{1/c^2}}$ and $h_{00}^{\ton{1/c^4}}$ denote the 1pN and second post--Newtonian (2pN) parts of $h_{00}$, respectively; both of them are needed to keep the Lagrangian to the 1pN level. To this aim,  it is meant that only $h_{00}^{\ton{1/c^2}}$ enters  the third and fourth  terms of \rfr{LpN}.

\textcolor{black}{Strictly speaking, the Lagrange equations in the form of \rfrs{dadt}{detadt} return either the osculating elements if $\mathfrak{R}$ depends only on the position $\boldsymbol{r}$ of the test particle, or the nonosculating, contact elements, to the first\textcolor{black}{\footnote{\textcolor{black}{In fact, when a velocity--dependent disturbing function is present, the Lagrange planetary equations for the contact elements are calculated with the replacement $\mathfrak{R}\rightarrow\mathfrak{R} + \ton{1/2}\ton{\partial\mathfrak{R}/\partial\boldsymbol{v}}^2$ \citep{1991ercm.book.....B,2011rcms.book.....K}\textcolor{black}{, where $\mathfrak{R}:=\mathcal{L}^\mathrm{pK}$}. The quadratic term yields also effects of the second order in the perturbation.}}} order in the perturbation, if $\mathfrak{R}$ does depend also on the velocity $\boldsymbol{v}$ \citep{1991ercm.book.....B,2011rcms.book.....K}. The consequences of this fact, often overlooked, will be treated in detail in Section \ref{sec:mike}. }

The orbital shifts experienced by the test particle during the flyby can be explicitly worked out by integrating the right hand sides of \rfrs{dadt}{detadt}, calculated onto the unperturbed Keplerian hyperbola, by means of \rfrs{sinhH}{coshH} and \rfrs{dtdf}{slrhyp}
from $f_\mathrm{min}$ to $f_\mathrm{max}$. As it will be shown, while for the 1pN LT and the Newtonian $J_2$ terms one can analytically work out shifts covering the whole motion by assuming
\eqi
\left| f_\mathrm{min}\right| = f_\mathrm{max} = f_\infty,\lb{diver}
\eqf
it is not possible for the 1pN gravitoelectric perturbations since they diverge when calculated with \rfr{diver}; however, analytical expressions valid for restricted ranges of values of $f$ including the passage at the pericentre can be obtained.
\section{The 1pN gravitoelectric shifts}\lb{sec:GE}
The 1pN gravitoelectric disturbing function, due solely to $M$, can be extracted from \rfr{LpN} by neglecting the last off--diagonal term and using \rfr{g00} and \rfr{gij} calculated for $J_2\rightarrow 0$.  It turns out to be
\eqi
\mathfrak{R}_\mathrm{GE} = \rp{r^2 v^4  + 12 \upmu r v^2 - 4\upmu^2  }{8 c^2 r^2}.\lb{RGE}
\eqf

By inserting \textcolor{black}{\rfr{RGE}} in \rfrs{dadt}{detadt} and integrating their right hand sides by means of \rfr{dtdf} within the range
\eqi
f_\mathrm{min}\leq f\leq f_\mathrm{max},\lb{kaz1}
\eqf
with
\eqi
-f_\infty < f_\mathrm{min} <0,
\eqf
and
\eqi
0  < f_\mathrm{max} < f_\infty,\lb{kaz2}
\eqf
one gets
\begin{align}
\Delta a^\mathrm{GE}\ton{f_\mathrm{min},f_\mathrm{max}} \lb{DaGE}& = 0,\acap
\Delta e^\mathrm{GE}\ton{f_\mathrm{min},f_\mathrm{max}} \lb{DeGE}& = 0,\acap
\Delta I^\mathrm{GE}\ton{f_\mathrm{min},f_\mathrm{max}} \lb{DIGE}& = 0,\acap
\Delta \Omega^\mathrm{GE}\ton{f_\mathrm{min},f_\mathrm{max}} \lb{DOGE}& = 0,\acap
\Delta \omega^\mathrm{GE}\ton{f_\mathrm{min},f_\mathrm{max}} \nonumber & = \rp{2\upmu}{c^2 a e^2}\ton{
-\rp{3\ton{1 + e^2}\Delta f}{e^2 - 1} - \rp{2\ton{3 + e^2}}{\sqrt{e^2 - 1}}\grf{\mathrm{arctanh}\qua{\rp{\ton{e - 1}\tan\ton{\rp{f_\mathrm{max}}{2}}}{\sqrt{e^2 - 1}}} - \mathrm{arctanh}\qua{\rp{\ton{e - 1}\tan\ton{\rp{f_\mathrm{min}}{2}}}{\sqrt{e^2 - 1}}}}\right.\acap
\lb{DoGE}&\left. - \rp{e\sin f_\mathrm{max}}{1 + e\cos f_\mathrm{max}} +  \rp{e\sin f_\mathrm{min}}{1 + e\cos f_\mathrm{min}}
},\acap
\textcolor{black}{\Delta \eta^\mathrm{GE}\ton{f_\mathrm{min},f_\mathrm{max}}} \nonumber & \textcolor{black}{=}  \textcolor{black}{\rp{\upmu}{2c^2 ae^2}}\ton{
\textcolor{black}{2\ton{12 - 11 e^2}\grf{\mathrm{arctanh}\qua{\rp{\ton{e - 1}\tan\ton{\rp{f_\mathrm{max}}{2}}}{\sqrt{e^2 - 1}}} - \mathrm{arctanh}\qua{\rp{\ton{e - 1}\tan\ton{\rp{f_\mathrm{min}}{2}}}{\sqrt{e^2 - 1}}}}}  \right.\acap
\lb{DetaGE}&\left. \textcolor{black}{-  \sqrt{e^2 - 1}\grf{12\Delta f + e\ton{e^2 - 4}\qua{\rp{\sin f_\mathrm{max}}{1 + e\cos f_\mathrm{max}} - \rp{\sin f_\mathrm{min}}{1 + e\cos f_\mathrm{min}}} }}
},
\end{align}
where
\eqi
\Delta f:= f_\mathrm{max} - f_\mathrm{min}.
\eqf
As anticipated in Section \ref{sec:calc}, \rfrs{DoGE}{DetaGE} turn out to be singular for
\begin{align}
f_\mathrm{min} & = -f_\infty,\acap
f_\mathrm{max} & = f_\infty.
\end{align}
It may happen that observations are collected during a larger time interval before the passage at the point of closest approach than after it, or vice versa; thus, the condition
\eqi
\left|f_\mathrm{min}\right| \neq \left|f_\mathrm{max}\right|\lb{diff}
\eqf
should be generally allowed. If, instead, data are taken during identical finite time spans before and after the flyby, it is
\eqi
f_\mathrm{min} = -f_\mathrm{max},\lb{fminmax}
\eqf
so that \rfrs{DoGE}{DetaGE} become
\begin{align}
\Delta\omega^\mathrm{GE}\ton{f_\mathrm{max}}\lb{DperiGE}& = -\rp{4\upmu}{c^2 a e^2}\grf{\rp{3\ton{1 + e^2}f_\mathrm{max}}{e^2 - 1}
+ \rp{2\ton{3 + e^2}\mathrm{arctanh}\qua{\rp{\ton{e - 1}\tan\ton{\rp{f_\mathrm{max}}{2}}}{\sqrt{e^2 - 1}}}}{\sqrt{e^2 - 1}} + \rp{e\sin f_\mathrm{max}}{1 + e\cos f_\mathrm{max}}
}, \acap
%
\textcolor{black}{\Delta \eta^\mathrm{GE}\ton{f_\mathrm{max}}} \lb{DetGE}& \textcolor{black}{=  \rp{\upmu}{c^2 ae^2}\grf{
2\ton{12 - 11 e^2}\mathrm{arctanh}\qua{\rp{\ton{e - 1}\tan\ton{\rp{f_\mathrm{max}}{2}}}{\sqrt{e^2 - 1}}} -  \sqrt{e^2 - 1}\qua{12 f_\mathrm{max} + \rp{e\ton{e^2 - 4}\sin f_\mathrm{max}}{1 + e\cos f_\mathrm{max}}
}}}
\end{align}
Expressions valid for short time intervals symmetric with respect to the flyby can be obtained by expanding \rfrs{DperiGE}{DetGE} in powers of $f_\mathrm{max}$, assumed close to zero; indeed, $f = 0$ corresponds just to the passage at pericentre. Thus, one obtains
\begin{align}
\Delta\omega_\mathrm{p}^\mathrm{GE}\ton{f_\mathrm{max}} \lb{bibi}&\simeq -\rp{4\upmu\ton{2 + e}}{c^2 a e\ton{e - 1}}f_\mathrm{max} + \mathcal{O}\ton{f_\mathrm{max}^2}, \acap
\textcolor{black}{\Delta\eta_\mathrm{p}^\mathrm{GE}\ton{f_\mathrm{max}}} \lb{bibo}& \textcolor{black}{\simeq \rp{\upmu \ton{8 + 3 e -10 e^2 - e^3}}{c^2 a e\sqrt{e^2 - 1}}f_\mathrm{max} + \mathcal{O}\ton{f_\mathrm{max}^2}.}
\end{align}

From \rfr{vinft} and \rfr{DaGE}, it can be straightforwardly inferred that $v_\infty$ is not changed by the 1pN gravitoelectric acceleration.
\section{The 1pN gravitomagnetic Lense--Thirring shifts}\lb{sec:LT}
The 1pN gravitomagnetic disturbing function, arising from the last term in \rfr{LpN} calculated with \rfr{h0i}, turns out to be
\eqi
\mathfrak{R}_\mathrm{LT} =  -\rp{2G J}{c^2 r^3}\ton{\boldsymbol{\hat{J}}\boldsymbol{\times}\boldsymbol{r}}\boldsymbol{\cdot}\boldsymbol{v}.\lb{RLT}
\eqf

Integrating \rfrs{dadt}{detadt}, calculated with \rfr{RLT}, by means of \rfrs{Df}{dtdf} finally yields
\begin{align}
\Delta a_\infty^\mathrm{LT} \lb{DaLT}& = 0,\acap
\Delta e_\infty^\mathrm{LT} \lb{DeLT}& = 0,\acap
\Delta I_\infty^\mathrm{LT} \lb{DILT}& = -\rp{4 G J \qua{\mathrm{arcsec}\ton{-e} + \sqrt{e^2 -1}}\texttt{Jl}}{c^2 \nk a^3\ton{e^2 - 1}^{3/2}},\acap
\Delta \Omega_\infty^\mathrm{LT} \lb{DOLT}& =  -\rp{4 G J\csc I\qua{\mathrm{arcsec}\ton{-e} + \sqrt{e^2 -1}}\texttt{Jm}}{c^2 a^3 \nk\ton{e^2 - 1}^{3/2}},\acap
\Delta \omega_\infty^\mathrm{LT} \lb{DoLT}& =  \rp{4 G J\grf{e^2\cot I\qua{\mathrm{arcsec}\ton{-e} + \sqrt{e^2 -1}}\texttt{Jm} + \qua{5 e^2 \mathrm{arcsec}\ton{-e} + \ton{3 + 2 e^2} \sqrt{e^2 -1}}\texttt{Jh}}}{c^2 a^3 \nk e^2\ton{e^2 - 1}^{3/2}},\acap
\textcolor{black}{\Delta \eta_\infty^\mathrm{LT}} \lb{DetaLT}& =  \textcolor{black}{-\rp{12 G J\sqrt{e^2 - 1}\texttt{Jh}}{c^2 a^3 \nk e^2}},
\end{align}
where
\begin{align}
\texttt{Jl} \lb{Jl}& := \boldsymbol{\hat{J}}\boldsymbol{\cdot}\boldsymbol{\hat{l}} = \hat{J}_x\cos\Omega + \hat{J}_y\sin\Omega, \acap
\texttt{Jm} \lb{Jm}& := \boldsymbol{\hat{J}}\boldsymbol{\cdot}\boldsymbol{\hat{m}} = \cos I\ton{-\hat{J}_x\sin\Omega + \hat{J}_y\cos\Omega} + \hat{J}_z\sin I, \acap
\texttt{Jl} \lb{Jh}& := \boldsymbol{\hat{J}}\boldsymbol{\cdot}\boldsymbol{\hat{h}} = \sin I\ton{\hat{J}_x\sin\Omega - \hat{J}_y\cos\Omega} + \hat{J}_z\cos I.
\end{align}
\Rfrs{DaLT}{DetaLT}, which cover the full motion of the test particle, retain a general validity since they hold for arbitrary orientations in space of both the orbit and the primary's spin axis.

From \rfrs{DaLT}{DetaLT} and \rfrs{Jl}{Jh} it turns out that the inclination and the node stay constant for equatorial orbits, characterized by
\begin{align}
\texttt{Jh}&=\pm 1,\acap
\texttt{Jl}&=\texttt{Jm}=0,
\end{align}
while the pericentre and the mean anomaly at epoch undergo nonvanishing net shifts. Instead, for polar orbits ($\texttt{Jh}=0$), the inclination, the node and the pericentre are, in general, shifted.

From \rfr{vinft} and \rfr{DaLT}, it can be straightforwardly inferred that $v_\infty$ is not changed by the LT acceleration.
\section{The Newtonian $J_2$ shifts}\lb{sec:J2}
The Newtonian disturbing function due to the primary's oblateness, obtained from the $J_2$--driven pK component of \rfr{LNt} calculated with \rfr{Upot},  turns out to be
\eqi
\mathfrak{R}_{J_2} = \rp{\upmu J_2 R^2_\mathrm{e}\qua{1 - 3 \ton{\boldsymbol{\hat{J}}\boldsymbol\cdot\boldsymbol{\hat{r}}}^2}}{2r^3}.\lb{RJ2}
\eqf
The resulting orbital shifts, integrated according to \rfrs{dadt}{detadt} and \rfrs{Df}{dtdf}, are
\begin{align}
\Delta a_\infty^{J_2} \lb{DaJ2}& = 0,\acap
\Delta e_\infty^{J_2} \lb{DeJ2}& = \rp{J_2 R_\mathrm{e}^2\sqrt{e^2 - 1}}{a^2 e^3}\sum_{i=1}^6 \mathcal{E}^{J_2}_{\infty,i}\widehat{T}_i,\acap
\Delta I_\infty^{J_2} \lb{DIJ2}& = \rp{J_2 R_\mathrm{e}^2}{a^2 e^2\ton{e^2 - 1}^2}\sum_{i=1}^6 \mathcal{I}^{J_2}_{\infty,i}\widehat{T}_i, \acap
\Delta \Omega_\infty^{J_2} \lb{DOJ2}& = \rp{J_2 R_\mathrm{e}^2\csc I}{a^2 e^2\ton{e^2 - 1}^2}\sum_{i=1}^6 \mathcal{N}^{J_2}_{\infty,i}\widehat{T}_i,\acap
\Delta \omega_\infty^{J_2} \lb{DoJ2} & = \rp{J_2 R_\mathrm{e}^2}{2 a^2 e^4\ton{e^2 - 1}^2}\sum_{i=1}^6 \mathcal{G}^{J_2}_{\infty,i}\widehat{T}_i, \acap
\Delta \eta_\infty^{J_2} \lb{DetaJ2}& = \rp{3J_2 R_\mathrm{e}^2}{2 a^2 e^4}\sum_{i=1}^6 \mathcal{H}^{J_2}_{\infty,i}\widehat{T}_i,
\end{align}
where
\begin{align}
\widehat{T}_1 \lb{T1} &: =1,\acap
\widehat{T}_2 \lb{T2} &: = \texttt{Jl}^2 + \texttt{Jm}^2 ,\acap
\widehat{T}_3 \lb{T3} &: = \texttt{Jl}^2 - \texttt{Jm}^2,\acap
\widehat{T}_4 \lb{T4} &: = \texttt{Jh}\,\texttt{Jl},\acap
\widehat{T}_5 \lb{T5} &: = \texttt{Jh}\,\texttt{Jm},\acap
\widehat{T}_6 \lb{T6} &: = \texttt{Jl}\,\texttt{Jm},
\end{align}
and
\begin{align}
\mathcal{E}^{J_2}_{\infty,1} \lb{EJ2infty1}&:= 0,\acap
\mathcal{E}^{J_2}_{\infty,2} \lb{EJ2infty2}&:= 0,\acap
\mathcal{E}^{J_2}_{\infty,3} \lb{EJ2infty3}&:= \sin 2\omega,\acap
\mathcal{E}^{J_2}_{\infty,4} \lb{EJ2infty4}&:= 0,\acap
\mathcal{E}^{J_2}_{\infty,5} \lb{EJ2infty5}&:= 0,\acap
\mathcal{E}^{J_2}_{\infty,6} \lb{EJ2infty6}&:= -2\mathcal{E}^{J_2}_{\infty,3}\cot 2\omega,\acap
\mathcal{I}^{J_2}_{\infty,1} \lb{IJ2infty1}&:= 0,\acap
\mathcal{I}^{J_2}_{\infty,2} \lb{IJ2infty2}&:= 0,\acap
\mathcal{I}^{J_2}_{\infty,3} \lb{IJ2infty3}&:= 0,\acap
\mathcal{I}^{J_2}_{\infty,4} \lb{IJ2infty4}&:= -3 e^2 \mathrm{arcsec}\ton{-e} + \sqrt{e^2 - 1}\qua{-3 e^2 - \ton{e^2 - 1}\cos 2\omega},\acap
\mathcal{I}^{J_2}_{\infty,5} \lb{IJ2infty5}&:= -\ton{e^2 - 1}^{3/2}\sin 2\omega,\acap
\mathcal{I}^{J_2}_{\infty,6} \lb{IJ2infty6}&:= 0,\acap
\mathcal{N}^{J_2}_{\infty,1} \lb{NJ2infty1}&:= 0,\acap
\mathcal{N}^{J_2}_{\infty,2} \lb{NJ2infty2}&:= 0,\acap
\mathcal{N}^{J_2}_{\infty,3} \lb{NJ2infty3}&:= 0,\acap
\mathcal{N}^{J_2}_{\infty,4} \lb{NJ2infty4}&:= \mathcal{I}^{J_2}_{\infty,5},\acap
\mathcal{N}^{J_2}_{\infty,5} \lb{NJ2infty5}&:= -3 e^2 \mathrm{arcsec}\ton{-e} + \sqrt{e^2 - 1}\qua{-3 e^2 + \ton{e^2 - 1}\cos 2\omega},\acap
\mathcal{N}^{J_2}_{\infty,6} \lb{NJ2infty6}&:= 0,\acap
\mathcal{G}^{J_2}_{\infty,1} \lb{GJ2infty1}&:= 6 e^2 \qua{\sqrt{e^2 - 1}\ton{1 + e^2} + 2 e^2 \mathrm{arcsec}\ton{-e}}, \acap
\mathcal{G}^{J_2}_{\infty,2} \lb{GJ2infty2}&:= -\rp{3}{2}\mathcal{G}_{\infty,1}^{J_2}, \acap
\mathcal{G}^{J_2}_{\infty,3} \lb{GJ2infty3}&:= -3 \sqrt{e^2 - 1} \ton{2 - 3 e^2 + e^4} \cos 2\omega, \acap
\mathcal{G}^{J_2}_{\infty,4} \lb{GJ2infty4}&:= 2 e^2 \ton{e^2 - 1}^{3/2} \cot I \sin 2\omega, \acap
\mathcal{G}^{J_2}_{\infty,5} \lb{GJ2infty5}&:= 2 e^2 \grf{3 e^2 \mathrm{arcsec}\ton{-e} + \sqrt{e^2 - 1} \qua{3 e^2 - \ton{e^2 - 1} \cos 2\omega}} \cot I, \acap
\mathcal{G}^{J_2}_{\infty,6} \lb{GJ2infty6}&:= -6 \sqrt{e^2 - 1} \ton{2 - 3 e^2 + e^4} \sin 2\omega,\acap
\mathcal{H}^{J_2}_{\infty,1} \lb{HJ2infty1}&:= -2 e^2, \acap
\mathcal{H}^{J_2}_{\infty,2} \lb{HJ2infty2}&:= -\rp{3}{2}\mathcal{H}^{J_2}_{\infty,1}, \acap
\mathcal{H}^{J_2}_{\infty,3} \lb{HJ2infty3}&:= \ton{2 + e^2}\cos 2\omega, \acap
\mathcal{H}^{J_2}_{\infty,4} \lb{HJ2infty4}&:= 0, \acap
\mathcal{H}^{J_2}_{\infty,5} \lb{HJ2infty5}&:= 0, \acap
\mathcal{H}^{J_2}_{\infty,6} \lb{HJ2infty6}&:= 2\mathcal{H}^{J_2}_{\infty,3}\tan 2\omega.
\end{align}
Also \rfrs{DaJ2}{DetaJ2}, covering the whole motion, are valid for any spatial orientations of the primary's spin axis and the orbital plane.

From \rfrs{DaJ2}{HJ2infty6} and \rfrs{Jl}{Jh}  it turns out that, for equatorial orbits, the eccentricity, the inclination and the node stay constant, while the pericentre and the mean anomaly at epoch do generally vary. Instead, for polar orbits, only the inclination and the node remain unaffected.

From \rfr{vinft} and \rfr{DaJ2}, it can be straightforwardly inferred that $v_\infty$ is not changed by the primary's oblateness.
\section{\textcolor{black}{A subtlety about the results obtained: choosing the gauge of the Lagrange planetary equations}}\lb{sec:mike}
\textcolor{black}{If the disturbing function depends also on the velocity $\boldsymbol v$, as in the case of the general relativistic \rfr{RGE} and \rfr{RLT}, the Lagrange planetary equations as given by \rfrs{dadt}{detadt} actually provide, to the first order in the perturbation, the instantaneous rates of change of the so--called contact Keplerian orbital elements. They are not osculating, and parameterize confocal instantaneous conics which may be generally neither tangent nor coplanar to the actual trajectory. Thus,  the correct position of the test particle is returned at each instant of time, but not its velocity since $\boldsymbol{v}\neq{\boldsymbol{v}}_\mathrm{K}$. Non--osculating Keplerian orbital elements were  used also for studying the impact of, e.g., $J_2$ \citep{2004CeMDA..90..289G} and  the 1pN gravitoelectric field \citep{2022AdSpR..69..538G} on bound, quasi--elliptical orbits. For a thorough analysis of the subtle technicalities involved, see \citet{2004A&A...415.1187E,2005NYASA1065..346E,2005CeMDA..91...75E,2011rcms.book.....K}. The physical and geometrical meaning of the contact elements is not as straightforward as for the osculating ones, and  wrong interpretations\footnote{For example, it was believed for a long time that, under certain assumptions, a near--equatorial satellite would always keep up with the equator of its host planet, assumed oblate, experiencing just oscillations of the  inclination, without any secular trends \citep{1965AJ.....70....5G,1993CeMDA..57..359K}. \textcolor{black}{The issue was later clarified by \citet{2005NYASA1065..346E}.}} of  (mathematically correct) results actually obtained in terms of nonosculating orbital elements can be found in the literature, as pointed out in \citet{2004A&A...415.1187E}. Thus, it may be preferable to have expressions for the orbital shifts written in terms of the osculating Keplerian  elements also for velocity--dependent disturbing functions. Indeed, in general, it is not guaranteed that, in the case of a hyperbolic motion, any variations of the  osculating and contact elements  coincide  when they are integrated over some range for $f$.}

\textcolor{black}{It turns out that the difference between the instantaneous values of the contact elements $C^\mathrm{ct}_i$ and the osculating ones $C^\mathrm{os}_i$ is given by \citep[p.\,74,\,Equation\,(1.323)]{2011rcms.book.....K}}
\eqi
\textcolor{black}{C^\mathrm{ct}_i\ton{f} - C^\mathrm{os}_i\ton{f} = -\sum_{j=1}^6\grf{C^\mathrm{os}_i,\,C^\mathrm{os}_j}\derp{\boldsymbol{r}}{C_j^\mathrm{os}}\boldsymbol\cdot\derp{\mathfrak{R}}{\boldsymbol{v}} := \mathcal{Z}_i\ton{f},\,i=a,e,I,\Omega,\omega,\eta,}\lb{mikeq}
\eqf
\textcolor{black}{where $\grf{C^\mathrm{os}_i,\,C^\mathrm{os}_j}$ are the Poisson brackets\textcolor{black}{\footnote{\textcolor{black}{Since they are time--independent, they can be evaluated for the value of $f$ which makes the calculation easier.}}} for the ith and jth elements.}

\textcolor{black}{A straightforward calculation yields}
\begin{align}
\textcolor{black}{\mathcal{Z}_a\ton{f}} \lb{Za}& \textcolor{black}{= \overline{\mathcal{Z}}_a^x\derp{\mathfrak{R}}{v_x} + \overline{\mathcal{Z}}_a^y\derp{\mathfrak{R}}{v_y} + \overline{\mathcal{Z}}_a^z\derp{\mathfrak{R}}{v_z},} \acap
\textcolor{black}{\mathcal{Z}_e\ton{f}} \lb{Ze}& \textcolor{black}{= \overline{\mathcal{Z}}_e^x\derp{\mathfrak{R}}{v_x} + \overline{\mathcal{Z}}_e^y\derp{\mathfrak{R}}{v_y} + \overline{\mathcal{Z}}_e^z\derp{\mathfrak{R}}{v_z},} \acap
\textcolor{black}{\mathcal{Z}_I\ton{f}} \lb{ZI}& \textcolor{black}{= \overline{\mathcal{Z}}_I^x\derp{\mathfrak{R}}{v_x} + \overline{\mathcal{Z}}_I^y\derp{\mathfrak{R}}{v_y} + \overline{\mathcal{Z}}_I^z\derp{\mathfrak{R}}{v_z},} \acap
\textcolor{black}{\mathcal{Z}_\Omega\ton{f}} \lb{ZO}& \textcolor{black}{= \overline{\mathcal{Z}}_\Omega^x\derp{\mathfrak{R}}{v_x} + \overline{\mathcal{Z}}_\Omega^y\derp{\mathfrak{R}}{v_y} + \overline{\mathcal{Z}}_\Omega^z\derp{\mathfrak{R}}{v_z},} \acap
\textcolor{black}{\mathcal{Z}_\omega\ton{f}} \lb{Zo}& \textcolor{black}{= \overline{\mathcal{Z}}_\omega^x\derp{\mathfrak{R}}{v_x} + \overline{\mathcal{Z}}_\omega^y\derp{\mathfrak{R}}{v_y} + \overline{\mathcal{Z}}_\omega^z\derp{\mathfrak{R}}{v_z},} \acap
\textcolor{black}{\mathcal{Z}_\eta\ton{f}} \lb{Zeta}& \textcolor{black}{= \overline{\mathcal{Z}}_\eta^x\derp{\mathfrak{R}}{v_x} + \overline{\mathcal{Z}}_\eta^y\derp{\mathfrak{R}}{v_y} + \overline{\mathcal{Z}}_\eta^z\derp{\mathfrak{R}}{v_z},} \acap
\end{align}
\textcolor{black}{where}
\begin{align}
\textcolor{black}{\overline{\mathcal{Z}}_a^x\ton{f}} \lb{Zax}& \textcolor{black}{= 0,}\acap
\textcolor{black}{\overline{\mathcal{Z}}_a^y\ton{f}} & \textcolor{black}{= 0,}\acap
\textcolor{black}{\overline{\mathcal{Z}}_a^z\ton{f}} & \textcolor{black}{= 0,}\acap
\textcolor{black}{\overline{\mathcal{Z}}_e^x\ton{f}} & \textcolor{black}{= \rp{\ton{e^2 - 1}^{3/2}\ton{\cos\Omega \sin u +
  \cos I\sin\Omega \cos u}}{ae\nk\ton{1 + e\cos f}},}\acap
\textcolor{black}{\overline{\mathcal{Z}}_e^y\ton{f}} & \textcolor{black}{= \rp{\ton{e^2 - 1}^{3/2}\ton{-\cos I\cos\Omega \cos u  + \sin\Omega\sin u}}{ae\nk\ton{1 + e\cos f}},}\acap
\textcolor{black}{\overline{\mathcal{Z}}_e^z\ton{f}} & \textcolor{black}{= -\rp{\ton{e^2 - 1}^{3/2}\sin I\cos u}{ae\nk\ton{1 + e\cos f}},}\acap
\textcolor{black}{\overline{\mathcal{Z}}_I^x\ton{f}} & \textcolor{black}{= -\rp{\sqrt{e^2 - 1}\sin I\sin\Omega\cos u}{a\nk\ton{1 + e\cos f}},}\acap
\textcolor{black}{\overline{\mathcal{Z}}_I^y\ton{f}} & \textcolor{black}{= \rp{\sqrt{e^2 - 1}\sin I\cos\Omega\cos u}{a\nk\ton{1 + e\cos f}},}\acap
\textcolor{black}{\overline{\mathcal{Z}}_I^z\ton{f}} & \textcolor{black}{= -\rp{\sqrt{e^2 - 1}\cos I\cos u}{a\nk\ton{1 + e\cos f}},}\acap
\textcolor{black}{\overline{\mathcal{Z}}_\Omega^x\ton{f}} & \textcolor{black}{= -\rp{\sqrt{e^2 - 1}\sin\Omega\sin u}{a\nk\ton{1 + e\cos f}},}\acap
\textcolor{black}{\overline{\mathcal{Z}}_\Omega^y\ton{f}} & \textcolor{black}{= \rp{\sqrt{e^2 - 1}\cos\Omega\sin u}{a\nk\ton{1 + e\cos f}},}\acap
\textcolor{black}{\overline{\mathcal{Z}}_\Omega^z\ton{f}} & \textcolor{black}{= -\rp{\sqrt{e^2 - 1}\cot I\sin u}{a\nk\ton{1 + e\cos f}},}\acap
\textcolor{black}{\overline{\mathcal{Z}}_\omega^x\ton{f}} & \textcolor{black}{= \rp{\sqrt{e^2 - 1} \grf{\cos\Omega\qua{2 e + \ton{1 + e^2} \cos f} \cos u  - \cos I\sin\Omega\ton{e + \cos f}\sin u}}{ae\nk\ton{1 + e\cos f}},}\acap
\textcolor{black}{\overline{\mathcal{Z}}_\omega^y\ton{f}} & \textcolor{black}{= \rp{\sqrt{e^2 - 1}\grf{\cos I \cos\Omega \ton{e + \cos f}\sin u + \sin\Omega\qua{2 e + \ton{1 + e^2} \cos f} \cos u}}{ae\nk\ton{1 + e\cos f}},}\acap
\textcolor{black}{\overline{\mathcal{Z}}_\omega^z\ton{f}} & \textcolor{black}{= \rp{\sqrt{e^2 - 1}\csc I\qua{3 e + \cos f + 2 e^2 \cos f - \cos 2I\ton{e + \cos f}}
\sin u}{ae\nk\ton{1 + e\cos f}},}\acap
\textcolor{black}{\overline{\mathcal{Z}}_\eta^x\ton{f}} & \textcolor{black}{= -\rp{\ton{e^2 - 1}^2 \ton{\cos\Omega\cos u -
   \cos I \sin\Omega\sin u}\cos f}{ae\nk\ton{1 + e\cos f}},}\acap
\textcolor{black}{\overline{\mathcal{Z}}_\eta^y\ton{f}} & \textcolor{black}{= -\rp{\ton{e^2 - 1}^2 \ton{\cos I \cos\Omega \sin u + \sin\Omega\cos u}\cos f}{ae\nk\ton{1 + e\cos f}},}\acap
\textcolor{black}{\overline{\mathcal{Z}}_\eta^z\ton{f}} \lb{Zetaz}& \textcolor{black}{= -\rp{\ton{e^2 - 1}^2 \sin I\cos f\sin u}{ae\nk\ton{1 + e\cos f}}.}
\end{align}
\textcolor{black}{In \rfrs{Zax}{Zetaz},}
\eqi
\textcolor{black}{u:=\omega + f}
\eqf
\textcolor{black}{is the argument of latitude. In general, also the components of the gradient of $\mathfrak{R}$ with respect to $\boldsymbol{v}$ are time--dependent through $f$.}

\textcolor{black}{The next step is  calculating the difference of the integrated shifts of the contact and osculating orbital elements.  To this aim, by taking the time derivative of both members of \rfr{mikeq}, one gets}
\eqi
\textcolor{black}{\dert{C^\mathrm{ct}_i\ton{f}}{t} - \dert{C^\mathrm{os}_i\ton{f}}{t} =\derp{\mathcal{Z}_i\ton{f}}{f}\dert{f}{t},\,i=a,e,I,\Omega,\omega,\eta.} \lb{diffk}
\eqf
\textcolor{black}{The analytical expressions for $\partial\mathcal{Z}_i\ton{f}/\partial f,\,i=a,e,I,\Omega,\omega,\eta$, obtainable straightforwardly from \rfrs{Za}{Zetaz}, are too cumbersome to be explicitly displayed. Integrating both members of \rfr{diffk} from $f_\mathrm{min}$ to $f_\mathrm{max}$ finally yields}
\eqi
\textcolor{black}{\Delta C^\mathrm{os}_i\ton{f_\mathrm{min},f_\mathrm{max}} =  \Delta C^\mathrm{ct}_i\ton{f_\mathrm{min},f_\mathrm{max}} + \Xi_i\ton{f_\mathrm{min},f_\mathrm{max}},\,i=a,e,I,\Omega,\omega,\eta,}
\eqf
\textcolor{black}{where}
\eqi
\textcolor{black}{\Xi_i\ton{f_\mathrm{min},f_\mathrm{max}} : = -\int_{f_\mathrm{min}}^{f_\mathrm{max}}\derp{\mathcal{Z}_i\ton{f}}{f}\mathrm{d}f,\,i=a,e,I,\Omega,\omega,\eta}.\lb{Csi}
\eqf
\textcolor{black}{In the case of the general relativistic disturbing functions of \rfr{RGE} and \rfr{RLT}, the integrated shifts of the contact elements $\Delta C_i^\mathrm{ct}\ton{f_\mathrm{min},f_\mathrm{max}},\,i=a,e,I,\Omega,\omega,\eta$ are given, to the 1pN order, by \rfrs{DaGE}{DetaGE} and \rfrs{DaLT}{DetaLT}, respectively.}

\textcolor{black}{From the above considerations, it turns out that the oblateness--driven classical orbital shifts of \rfrs{DaJ2}{DetaJ2} are to be intended as written in terms of the osculating elements. Indeed, the disturbing function of \rfr{RJ2} depends only on the position vector $\boldsymbol{r}$. In this case, the contact and the osculating elements coincide, as per \rfr{mikeq}.}

\textcolor{black}{For elliptical motions perturbed by \rfr{RGE} and \rfr{RLT}, it can be shown that $\Xi_i\ton{0,2\uppi} = 0,\,i = a,e,I,\Omega,\omega,\eta,$ when \rfr{Csi} is integrated from $0$ to $2\uppi$.}
\subsection{\textcolor{black}{The 1pN gravitoelectric corrections to the shifts of the contact Keplerian orbital elements}}\lb{cont:GE}
\textcolor{black}{The corrections $\Xi_i^\mathrm{GE}\ton{f_\mathrm{min},f_\mathrm{max}},\,i=a,e,I,\Omega,\omega,\eta$ to the 1pN gravitoelectric integrated shifts of the contact Keplerian orbital elements are obtained in the following way.}

\textcolor{black}{The calculation of the gradient of \rfr{RGE} with respect to $\boldsymbol{v}$ returns}
\begin{align}
\textcolor{black}{\derp{\mathfrak{R}_\mathrm{GE}}{v_x}} \lb{dRGEdvx}& \textcolor{black}{= \rp{\upmu\nk\ton{7 + e^2 +
8 e \cos f} \qua{\cos\Omega \ton{e \sin\omega + \sin u} + \cos I\sin\Omega \ton{e \cos\omega + \cos u}}}{2c^2\ton{e^2 - 1}^{3/2}}, } \acap
\textcolor{black}{\derp{\mathfrak{R}_\mathrm{GE}}{v_y}} \lb{dRGEdvy}& \textcolor{black}{= -\rp{\upmu\nk\ton{7 + e^2 +
8 e \cos f} \qua{\cos I\cos\Omega \ton{e \cos\omega + \cos u} - \sin\Omega\ton{e \sin\omega + \sin u}}}{2c^2\ton{e^2 - 1}^{3/2}}, } \acap
\textcolor{black}{\derp{\mathfrak{R}_\mathrm{GE}}{v_z}} \lb{dRGEdvz}& \textcolor{black}{= -\rp{\upmu\nk\sin I\ton{7 + e^2 + 8 e \cos f} \ton{e \cos\omega +
\cos u}}{2c^2\ton{e^2 - 1}^{3/2}}. }
\end{align}

\textcolor{black}{The functions $\mathcal{Z}^\mathrm{GE}_i\ton{f},\,i=a,e,I,\Omega,\omega,\eta$ entering \rfr{mikeq} can be calculated by inserting \Rfrs{dRGEdvx}{dRGEdvz} in \rfrs{Zax}{Zetaz}. One gets}
\begin{align}
\textcolor{black}{\mathcal{Z}^\mathrm{GE}_a\ton{f}} \lb{ZaGE}& \textcolor{black}{= 0,}\acap
\textcolor{black}{\mathcal{Z}^\mathrm{GE}_e\ton{f}} \lb{ZeGE}& \textcolor{black}{= -\rp{\upmu\ton{7 + e^2 + 8 e \cos f}}{2c^2 a e},}\acap
\textcolor{black}{\mathcal{Z}^\mathrm{GE}_I\ton{f}} \lb{ZIGE}& \textcolor{black}{= 0,}\acap
\textcolor{black}{\mathcal{Z}^\mathrm{GE}_\Omega\ton{f}} \lb{ZOGE}& \textcolor{black}{= 0,}\acap
\textcolor{black}{\mathcal{Z}^\mathrm{GE}_\omega\ton{f}} \lb{ZoGE}& \textcolor{black}{= \rp{\upmu\qua{6 e \ton{3 + e^2} + \ton{7 + 24 e^2 + e^4} \cos f + 4 e\ton{1 + e^2} \cos 2f}\sin f}{2c^2 a\ton{e^2 - 1} \ton{1 + e\cos f}^2},}\acap
\textcolor{black}{\mathcal{Z}^\mathrm{GE}_\eta\ton{f}} \lb{ZetaGE}& \textcolor{black}{= -\rp{\upmu\sqrt{e^2 - 1}\ton{7 + e^2 + 8 e \cos f}\sin 2 f}{4c^2 a\ton{1 + e\cos f}^2}.}
\end{align}
\textcolor{black}{According to \rfrs{ZaGE}{ZetaGE}, the instantaneous osculating eccentricity, argument of pericenter and mean anomaly at epoch are generally different from their contact counterparts, while the osculating and contact semimajor axis, inclination and longitude of the ascending node always coincide.}

\textcolor{black}{By plugging \rfrs{ZaGE}{ZetaGE} in \rfr{Csi}, the explicit expressions of $\Xi_i^\mathrm{GE}\ton{f_\mathrm{min},f_\mathrm{max}},\,i=a,e,I,\Omega,\omega,\eta$ can be obtained. They are}
\begin{align}
\textcolor{black}{\Xi_a^\mathrm{GE}\ton{f_\mathrm{min},f_\mathrm{max}}} \lb{Xia}& \textcolor{black}{= 0,}\acap
\textcolor{black}{\Xi_e^\mathrm{GE}\ton{f_\mathrm{min},f_\mathrm{max}}} \lb{Xie}& \textcolor{black}{=\rp{4\upmu\ton{\cos f_\mathrm{min} - \cos f_\mathrm{max}} }{c^2 a},}\acap
\textcolor{black}{\Xi_I^\mathrm{GE}\ton{f_\mathrm{min},f_\mathrm{max}}} \lb{XiI}& \textcolor{black}{= 0,}\acap
\textcolor{black}{\Xi_\Omega^\mathrm{GE}\ton{f_\mathrm{min},f_\mathrm{max}}} \lb{XiO}& \textcolor{black}{= 0,}\acap
\textcolor{black}{\Xi_\omega^\mathrm{GE}\ton{f_\mathrm{min},f_\mathrm{max}}} \nonumber & \textcolor{black}{=\rp{\upmu}{2c^2 a\ton{e^2 - 1}}}\grf{\textcolor{black}{\rp{ \qua{6 e \ton{3 + e^2} + \ton{7 + 24 e^2 + e^4} \cos f_\mathrm{min} + 4 e\ton{1 + e^2} \cos 2f_\mathrm{min}} \sin f_{min}  }{\ton{1 + e\cos f_\mathrm{min}}^2} } \right.\acap
\lb{Xio}&\left. \textcolor{black}{-\rp{\qua{6 e \ton{3 + e^2} + \ton{7 + 24 e^2 + e^4} \cos f_\mathrm{max} + 4 e\ton{1 + e^2} \cos 2f_\mathrm{max}} \sin f_{max}  }{\ton{1 + e\cos f_\mathrm{max}}^2}}}, \acap
\textcolor{black}{\Xi_\eta^\mathrm{GE}\ton{f_\mathrm{min},f_\mathrm{max}}} \lb{Xiet}& \textcolor{black}{=\rp{\upmu\sqrt{e^2 - 1}}{2c^2 a}\qua{\rp{\ton{7 + e^2 + 8 e \cos f_\mathrm{max}} \sin 2f_\mathrm{max}}{2\ton{1 + e\cos f_\mathrm{max}}^2} - \rp{\ton{7 + e^2 + 8 e \cos f_\mathrm{min}} \sin 2f_\mathrm{min}}{2\ton{1 + e\cos f_\mathrm{min}}^2}}.}
\end{align}
\textcolor{black}{Also \rfrs{Xia}{Xiet}, as \rfrs{DaGE}{DetaGE},  hold only for the range of values of $f$ given by \rfrs{kaz1}{kaz2}.}

\textcolor{black}{By using \rfr{fminmax}, \rfrs{Xie}{Xiet} reduce to}
\begin{align}
\textcolor{black}{\Xi_a^\mathrm{GE}\ton{f_\mathrm{max}}} \lb{xia}& \textcolor{black}{= 0,}\acap
\textcolor{black}{\Xi_e^\mathrm{GE}\ton{f_\mathrm{max}}} \lb{xie}& \textcolor{black}{= 0,}\acap
\textcolor{black}{\Xi_I^\mathrm{GE}\ton{f_\mathrm{max}}} \lb{xiI}& \textcolor{black}{= 0,}\acap
\textcolor{black}{\Xi_\Omega^\mathrm{GE}\ton{f_\mathrm{max}}} \lb{xiO}& \textcolor{black}{= 0,}\acap
\textcolor{black}{\Xi_\omega^\mathrm{GE}\ton{f_\mathrm{max}}} \lb{xio} & \textcolor{black}{= -\rp{\upmu\qua{6 e \ton{3 + e^2} + \ton{7 + 24 e^2 + e^4} \cos f_\mathrm{max} + 4 e\ton{1 + e^2} \cos 2f_\mathrm{max}} \sin f_{max}}{c^2 a\ton{e^2 - 1}\ton{1 + e\cos f_\mathrm{max}}^2},} \acap
\textcolor{black}{\Xi_\eta^\mathrm{GE}\ton{f_\mathrm{max}}} \lb{xiet} & = \textcolor{black}{\rp{\upmu\sqrt{e^2 - 1}\ton{7 + e^2 + 8 e \cos f_\mathrm{max}} \sin 2f_\mathrm{max}}{2c^2 a\ton{1 + e\cos f_\mathrm{max}}^2}.}
\end{align}

\textcolor{black}{By expanding \rfrs{xio}{xiet} in powers of $f_\mathrm{max}$ around 0, as done for \rfrs{bibi}{bibo}, yields}
\begin{align}
\textcolor{black}{\Xi_\omega^\mathrm{GE}\ton{f_\mathrm{max}}} \lb{xino} & \textcolor{black}{\simeq \rp{\upmu \ton{7 + e}}{c^2 a\ton{1 - e}} f_\mathrm{max} + \mathcal{O}\ton{f^2_\mathrm{max}},} \acap
\textcolor{black}{\Xi_\eta^\mathrm{GE}\ton{f_\mathrm{max}}} \lb{xinet} & \textcolor{black}{\simeq  \rp{\upmu\ton{7 + e}\sqrt{e^2 - 1}}{c^2 a\ton{e + 1}} f_\mathrm{max} + \mathcal{O}\ton{f^2_\mathrm{max}}.}
\end{align}
\subsection{\textcolor{black}{The Lense--Thirring corrections to the shifts of the contact Keplerian orbital elements}}\lb{cont:LT}
\textcolor{black}{Explicit expressions for the corrections $\Xi_i^\mathrm{LT}\ton{f_\mathrm{min},f_\mathrm{max}},\,i=a,e,I,\Omega,\omega,\eta$ to the LT integrated shifts of the contact Keplerian orbital elements can be obtained as follows.}

\textcolor{black}{Calculating the gradient of \rfr{RLT} with respect to $\boldsymbol{v}$ yields}
\begin{align}
\textcolor{black}{\derp{\mathfrak{R}_\mathrm{LT}}{v_x}} \lb{dRLTdvx}& \textcolor{black}{= \rp{2GJ\ton{1 + e\cos f}^2\qua{\ton{\hat{J}_z \cos I \cos\Omega - \hat{J}_y \sin I} \sin u + \hat{J}_z \sin\Omega\cos u }}{c^2a^2\ton{e^2 - 1}^2}}, \acap
\textcolor{black}{\derp{\mathfrak{R}_\mathrm{LT}}{v_y}} \lb{dRLTdvy}& \textcolor{black}{= -\rp{2GJ\ton{1 + e\cos f}^2\qua{\hat{J}_z \cos\Omega\cos u -
 \ton{\hat{J}_x \sin I + \hat{J}_z \cos I \sin\Omega}\sin u}}{c^2a^2\ton{e^2 - 1}^2}}, \acap
\textcolor{black}{\derp{\mathfrak{R}_\mathrm{LT}}{v_z}} \lb{dRLTdvz}& \textcolor{black}{= \rp{2GJ\ton{1 + e\cos f}^2\qua{\ton{\hat{J}_y \cos\Omega -
\hat{J}_x \sin\Omega}\cos u - \cos I\ton{\hat{J}_x \cos\Omega + \hat{J}_y \sin\Omega}\sin u}}{c^2a^2\ton{e^2 - 1}^2}.}
\end{align}

\textcolor{black}{\Rfrs{dRLTdvx}{dRLTdvz}, inserted in \rfrs{Zax}{Zetaz}, allow to calculate the functions $\mathcal{Z}^\mathrm{LT}_i\ton{f},\,i=a,e,I,\Omega,\omega,\eta$ entering \rfr{mikeq}. They turn out to be}
\begin{align}
\textcolor{black}{\mathcal{Z}^\mathrm{LT}_a\ton{f}} \lb{ZaLT}& \textcolor{black}{= 0,}\acap
\textcolor{black}{\mathcal{Z}^\mathrm{LT}_e\ton{f}} \lb{ZeLT}& \textcolor{black}{= \rp{2GJ\ton{1 + e\cos f}\texttt{Jh}}{c^2 a^3 \nk e\sqrt{e^2 - 1}},}\acap
\textcolor{black}{\mathcal{Z}^\mathrm{LT}_I\ton{f}} \lb{ZILT}& \textcolor{black}{= -\rp{2GJ\ton{1 + e\cos f}\cos u\ton{\texttt{Jm}\cos u - \texttt{Jl}\sin u}}{c^2 a^3 \nk\ton{e^2 - 1}^{3/2}},}\acap
\textcolor{black}{\mathcal{Z}^\mathrm{LT}_\Omega\ton{f}} \lb{ZOLT}& \textcolor{black}{= -\rp{2GJ\csc I\ton{1 + e\cos f}\sin u\ton{\texttt{Jm}\cos u - \texttt{Jl}\sin u}}{c^2 a^3 \nk\ton{e^2 - 1}^{3/2}},}\acap
\textcolor{black}{\mathcal{Z}^\mathrm{LT}_\omega\ton{f}} \lb{ZoLT}& \textcolor{black}{= \rp{2GJ\cot I\ton{1 + e\cos f}\sin u\ton{\texttt{Jm}\cos u - \texttt{Jl}\sin u}}{c^2 a^3 \nk\ton{e^2 - 1}^{3/2}},}\acap
\textcolor{black}{\mathcal{Z}^\mathrm{LT}_\eta\ton{f}} \lb{ZetaLT}& \textcolor{black}{= 0.}
\end{align}
\textcolor{black}{According to \rfrs{ZaLT}{ZetaLT}, the instantaneous osculating eccentricity, inclination, longitude of ascending node and argument of pericenter are generally different from their contact counterparts, while the osculating and contact semimajor axis and mean anomaly at epoch always coincide.}

\textcolor{black}{By means of \rfr{Csi}, calculated with \rfrs{ZaLT}{ZetaLT}, one straightforwardly obtains the explicit expressions of $\Xi_i^\mathrm{LT}\ton{f_\mathrm{min},f_\mathrm{max}},\,i=a,e,I,\Omega,\omega,\eta$ which, apart from the semimajor axis and the mean anomaly at epoch,  turn out to be generally nonvanishing functions valid for any values of $f_\mathrm{min}$ and $f_\mathrm{max}$. They are too cumbersome to be explicitly displayed for the general case of \rfr{diff}.
Instead, it is possible to write down manageable formulas if the condition of \rfr{fminmax} holds. The resulting nonvanishing shifts are}
\begin{align}
\textcolor{black}{\Xi_I^\mathrm{LT}\ton{f_\mathrm{max}}} \lb{kuz1}& \textcolor{black}{= \rp{2GJ\ton{1 + e\cos f_\mathrm{max}}\sin 2f_\mathrm{max}\ton{\texttt{Jl}\cos 2\omega + \texttt{Jm}\sin 2\omega}}{c^2 a^3\nk\ton{e^2 - 1}^{3/2}},}\acap
\textcolor{black}{\Xi_\Omega^\mathrm{LT}\ton{f_\mathrm{max}}} & \textcolor{black}{= \rp{2GJ\csc I\ton{1 + e\cos f_\mathrm{max}}\sin 2f_\mathrm{max}\ton{\texttt{Jl}\sin 2\omega - \texttt{Jm}\cos 2\omega}}{c^2 a^3\nk\ton{e^2 - 1}^{3/2}},}\acap
\textcolor{black}{\Xi_\omega^\mathrm{LT}\ton{f_\mathrm{max}}} \lb{kuz3}& \textcolor{black}{= \rp{2GJ\cot I\ton{1 + e\cos f_\mathrm{max}}\sin 2f_\mathrm{max}\ton{-\texttt{Jl}\sin 2\omega + \texttt{Jm}\cos 2\omega}}{c^2 a^3\nk\ton{e^2 - 1}^{3/2}}.}\acap
\end{align}
\textcolor{black}{\Rfrs{kuz1}{kuz3} vanish if}
\eqi
\textcolor{black}{f_\mathrm{max} = f_\infty.}
\eqf
\textcolor{black}{Thus, the LT total shifts of the osculating orbital elements accumulated over the entire path are equal just to those given by \rfrs{DaLT}{DetaLT}.}

\section{Numerical evaluations for some natural and artificial bodies}\lb{sec:num}
Here, the results obtained in the previous Sections are applied to some flybys occurred in our solar system.
\subsection{ʻOumuamua in the field of the Sun}\lb{sec:Oum}
The case of the interstellar, cigar--shaped  asteroid\footnote{In view of the remarkable non--gravitational acceleration exhibited by ʻOumuamua, not accompanied by typical cometary activity tracers, it was argued that it may be an artifact by some alien civilization \citep{2018ApJ...868L...1B}. Later, such a hypothesis was dismissed in favor of a conventional explanation based on known non--gravitational physics \citep{2023Natur.615..610B}.} 1I/2017 U1 (ʻOumuamua) \citep{2017Natur.552..378M}, which briefly visited the inner regions of out solar system in 2017 along an unbound trajectory, is considered here. Its orbital parameters are listed in Table \ref{Tab:Oum}.
\begin{table}[h]
\caption{Orbital parameters of the heliocentric trajectory of the interstellar asteroid  ʻOumuamua referred to the International Celestial Reference Frame (ICRF) at epoch J2000.0. retrieved from the HORIZONS WEB interface maintained by the Jet Propulsion Laboratory (JPL) for the epoch 23th November 2017. The distance of closest approach turns out to be $r_\mathrm{p}=0.38\,\mathrm{au} = 82\,R_\mathrm{e}^\odot$, while $f_\infty = 146.4\,\mathrm{deg} = 2.55\,\mathrm{rad}$.
}\lb{Tab:Oum}
\begin{center}
\begin{tabular}{|l|l|l|}
  \hline
Parameter  & Units & Numerical Value \\
\hline
$a$ & au & $-1.9$\\
$e$ & $-$ & $1.2$\\
$I$ & deg & $143.1$\\
$\Omega$ & deg & $35.7$\\
$\omega$ & deg & $257.8$\\
\hline
\end{tabular}
\end{center}
\end{table}
It may be interesting to calculate the size of the pK effects of gravitational origin treated in the previous Sections if only to get an idea of the
potential offered by this type of unusual objects
\citep{2024arXiv240706475J} whose number may increase in the future\footnote{Only one other object belonging to this class has been discovered so far: the very active interstellar comet 2I/Borisov \citep{2020NatAs...4...53G}.}. However, it should be made clear that using them as probes for tests of gravitational theories would be quite challenging non only because of the observational accuracy needed but also because of the heavy non--gravitational accelerations perturbing their motion \citep{2018Natur.559..223M}.

In Table \ref{Tab:Sun}, the relevant physical parameters of the Sun are listed.
\begin{table}[h]
\caption{Relevant physical parameters of the Sun \citep{1998MNRAS.297L..76P,2007CeMDA..98..155S,2012ApJ...750..135E,2017AJ....153..121P,2021MNRAS.506.2671M,2021AJ....161..105P}. R.A. $\alpha_\odot$ and decl. $\delta_\odot$ of the north pole of rotation are equatorial coordinates referred to  the International Celestial Reference Frame (ICRF) at epoch J2000.0.
}\lb{Tab:Sun}
\begin{center}
\begin{tabular}{|l|l|l|}
  \hline
Parameter  & Units & Numerical Value \\
\hline
$\upmu_\odot$ & $\times 10^{20}\,\textrm{m}^3/\textrm{s}^2$ & $1.32712440041279419$ \citep{2021AJ....161..105P}\\
$J_2^\odot$ & $\times 10^{-7}$ & $2.2$ \citep{2017AJ....153..121P,2021MNRAS.506.2671M}\\
$J_\odot$ & $\times 10^{41}\,\textrm{kg~m}^2$$/\textrm{s}$ & $1.90$ \citep{1998MNRAS.297L..76P}\\
$\alpha_\odot$ & deg & $286.13$ \citep{2007CeMDA..98..155S}\\
$\delta_\odot$ & deg & $63.87$ \citep{2007CeMDA..98..155S}\\
$R_\mathrm{e}^\odot$ & $\textrm{km}$ & $696342$ \citep{2012ApJ...750..135E}\\
\hline
\end{tabular}
\end{center}
\end{table}
The components of the Sun's spin axis, parameterized in terms of the right ascension (R.A.) $\alpha_\odot$ and declination (decl.) $\delta_\odot$ of its north pole of rotation, are
\begin{align}
\hat{J}_x^\odot \lb{Jxs}& = \cos\alpha_\odot\cos\delta_\odot,\acap
\hat{J}_y^\odot \lb{Jys}& = \sin\alpha_\odot\cos\delta_\odot,\acap
\hat{J}_z^\odot \lb{Jzs}& = \sin\delta_\odot;
\end{align}
they are needed to calculate \rfrs{DaLT}{DetaLT} and \rfrs{DaJ2}{DetaJ2}.

The nominal values of the gravitational pK orbital shifts of ʻOumuamua are displayed in  Table \ref{Tab:Oum_pK}.
\begin{table}[h]
\caption{Nominal values of the pK orbital shifts of ʻOumuamua calculated with the values of Table \ref{Tab:Sun} and Table \ref{Tab:Oum}; $f_\mathrm{max}$ entering the 1pN gravitoelectric shifts is assumed to be close to $0$, and its value has to be given in rad. Here, mas and $\mu$as stand for milliarcseconds and microarcseconds, respectively.
}\lb{Tab:Oum_pK}
\begin{center}
\begin{tabular}{|l|l|l|}
  \hline
Parameter  & Units & Numerical Value \\
\hline
$\Delta\omega^\mathrm{GE}$ & mas & $ \textcolor{black}{100.3}\,f_\mathrm{max}$\\
$\Delta\eta^\mathrm{GE}$ & mas & $ \textcolor{black}{3.4}\,f_\mathrm{max}$\\
$\Delta I_\infty^\mathrm{LT}$ & $\mu$as & $-0.1$\\
$\Delta \Omega_\infty^\mathrm{LT}$ & $\mu$as & $1.0$\\
$\Delta \omega_\infty^\mathrm{LT}$ & $\mu$as & $2.5$\\
$\Delta \eta_\infty^\mathrm{LT}$ & $\mu$as & $\textcolor{black}{-0.04}$\\
$\Delta e_\infty^{J_2}$ & - & $ -2\times 10^{-13} $\\
$\Delta I_\infty^{J_2}$ & $\mu$as & $-0.8$\\
$\Delta \Omega_\infty^{J_2}$ & $\mu$as & $8.4$\\
$\Delta \omega_\infty^{J_2}$ & $\mu$as & $2.8$\\
$\Delta \eta_\infty^{J_2}$ & $\mu$as & $0.4$\\
\hline
\end{tabular}
\end{center}
\end{table}
They are exceedingly small. Suffice it to say that the effects of the Sun's oblateness and angular momentum are at the microarcseconds ($\mu$as) level over the whole trajectory, while the 1pN gravitoelectric shifts, to be rescaled by $f_\mathrm{max}\gtrsim 0$ since they are valid just around the flyby, are of the order of less than a \textcolor{black}{hundred} milliarcseconds (mas).
\subsection{NEAR in the field of the Earth}\lb{sec:Near}
Here, the case of the spacecraft NEAR\footnote{It was a man--made robotic probe designed to study the near--Earth asteroid (433) Eros \citep{2002AcHA...15..210S}. Its mission profile included, among other things, a flyby of the Earth.} \citep{2002AcAau..51..491P} when it approached the Earth is treated.

Table \ref{Tab:Near} lists the orbital parameters of such a spacecraft for the flyby of the Earth occurred on 23th January 1998.
\begin{table}[h]
\caption{Orbital parameters of the geocentric trajectory of the probe NEAR referred to the International Celestial Reference Frame (ICRF) at epoch J2000.0. retrieved from the HORIZONS WEB interface maintained by the Jet Propulsion Laboratory (JPL) for the epoch 23th January 1998. The distance of closest approach turns out to be $r_\mathrm{p} = 6.90\times 10^3\,\mathrm{km} = 1.08\,R_\mathrm{e}^\oplus$, while $f_\infty = 123.4\,\mathrm{deg} = 2.15\,\mathrm{rad}$.
}\lb{Tab:Near}
\begin{center}
\begin{tabular}{|l|l|l|}
  \hline
Parameter  & Units & Numerical Value \\
\hline
$a$ & km & $-8.49\times 10^3$\\
$e$ & $-$ & $1.813$\\
$I$ & deg & $107.97$\\
$\Omega$ & deg & $88.2$\\
$\omega$ & deg & $145.1$\\
\hline
\end{tabular}
\end{center}
\end{table}
Table \ref{Tab:Near_pK} displays the nominal values for the pK orbital shifts experienced by NEAR; the relevant physical parameters of the Earth needed to calculate them were retrieved from \citet{iers10}.
\begin{table}[h]
\caption{Nominal values of the pK orbital shifts of the probe NEAR calculated with the values of Table \ref{Tab:Near}; $f_\mathrm{max}$ entering the 1pN gravitoelectric shifts is assumed to be close to $0$, and its value has to be given in rad. Here, mas and $\mu$as stand for milliarcseconds and microarcseconds, respectively.
}\lb{Tab:Near_pK}
\begin{center}
\begin{tabular}{|l|l|l|}
  \hline
Parameter  & Units & Numerical Value \\
\hline
$\Delta\omega^\mathrm{GE}$ & mas & $ \textcolor{black}{2.3}\,f_\mathrm{max}$\\
$\Delta\eta^\mathrm{GE}$ & mas & $ \textcolor{black}{0.5}\,f_\mathrm{max}$\\
$\Delta I_\infty^\mathrm{LT}$ & $\mu$as & $0$\\
$\Delta \Omega_\infty^\mathrm{LT}$ & $\mu$as & $7.7$\\
$\Delta \omega_\infty^\mathrm{LT}$ & $\mu$as & $12.2$\\
$\Delta \eta_\infty^\mathrm{LT}$ & $\mu$as & $\textcolor{black}{-3.1}$\\
$\Delta e_\infty^{J_2}$ & - & $ 0.0001 $\\
$\Delta I_\infty^{J_2}$ & $\mu$as & $-7\times 10^6$\\
$\Delta \Omega_\infty^{J_2}$ & $\mu$as & $7.9\times 10^7$\\
$\Delta \omega_\infty^{J_2}$ & $\mu$as & $-1.3\times 10^8$\\
$\Delta \eta_\infty^{J_2}$ & $\mu$as & $1.2\times 10^7$\\
\hline
\end{tabular}
\end{center}
\end{table}
It turns out that the nominal orbital shifts due to the Earth's first even zonal harmonic, being as large as $\simeq 10^6-10^8\,\mu\mathrm{as}$, neatly overwhelm the pN ones; suffice it to say that the LT displacements are as little as $\lesssim 10\,\mu\mathrm{as}$, while the gravitoelectric ones are \textcolor{black}{less than a few mas}.
However, the present--day relative uncertainty in determining $J_2^\oplus$ from several dedicated satellite missions is\footnote{Such an evaluation is based just on the formal, statistical errors $\sigma_{J_2^\oplus}$ of the most recent solutions for the Earth's gravity field.}
\eqi
\rp{\sigma_{J_2^\oplus}}{J_2^\oplus}\simeq 10^{-8},
\eqf
as it can be inferred by inspecting the latest Earth's gravity models retrievable from, e.g., the webpage\footnote{See \url{https://icgem.gfz-potsdam.de/tom\_longtime} on the Internet.} of the International Centre for Global Earth Models (ICGEM) maintained by the GeoForschungsZentrum (GFZ). Thus, the mismodelled classical shifts would be smaller than the nominal pN ones by about one order of magnitude (LT), or more (Schwarzschild). Also in the case of artificial probes like NEAR, the impact of the non--gravitational accelerations during flybys should be carefully investigated.

\textcolor{black}{Given the low altitude}
\eqi
\textcolor{black}{h_\mathrm{p} \simeq 532\,\mathrm{km}}
\eqf
\textcolor{black}{reached by NEAR at the perigee of its Earth's flyby, one may wonder what could be the impact of, say, the mismodeling $\sigma_{J^\oplus_4}$ of the second even zonal harmonic $J_4^\oplus$ of the geopotential. It can be naively evaluated by multiplying the nominal values of the shifts due to $J_2$ listed in Table \ref{Tab:Near_pK} by the following scaling factor
}
\eqi
\textcolor{black}{\upalpha_{J^\oplus_4}\simeq\ton{\rp{R}{a}}^2\rp{\sigma_{J_4^\oplus}}{J_2^\oplus}.}\lb{ska4}
\eqf
\textcolor{black}{The square of the ratio of the Earth's equatorial radius to the semimajor axis of NEAR, listed in Table \ref{Tab:Near}, is of the order of} \eqi
\textcolor{black}{\ton{\rp{R}{a}}^2\simeq 0.5.}
\eqf
\textcolor{black}{According to the most recent Earth's gravity models listed by ICGEM, the formal uncertainty in $J_4^\oplus$ is as little as}
\eqi
\textcolor{black}{\sigma_{J_4^\oplus}\simeq 10^{-13}.}
\eqf
\textcolor{black}{The nominal value of the first even zonal harmonic of our planet is}
\eqi
\textcolor{black}{J_2^\oplus\simeq 10^{-4}.}
\eqf
\textcolor{black}{Thus, the bias due to the imperfect knowledge of the Earth's octupole mass moment seems somewhat negligible with respect to the pN effects of interest, shown in Table \ref{Tab:Near_pK}, since
}
\eqi
\textcolor{black}{\upalpha_{J^\oplus_4}\simeq 10^{-10}.}
\eqf

\textcolor{black}{Finally, as a further source of potential systematic bias, it may be mentioned the coefficient of degree $\ell=2$ and order $m=2$ of the geopotential accounting for the Earth's dynamical triaxiality. Its impact on spacecraft's flybys was analytically investigated by, e.g., \citet{2001Icar..150..168R}.
}
\section{Summary and conclusions}\lb{Sec:end}
Analytical expressions of the variations of all the Keplerian orbital elements of an otherwise unperturbed hyperbolic trajectory are obtained in full generality for some known post--Keplerian perturbing accelerations of both Newtonian and post--Newtonian origin: that due to the primary's quadrupole mass moment, and, to the first post--Newtonian level, the general relativistic Schwarzschild and Lense--Thirring ones. \textcolor{black}{In the case of the classical perturbation, the resulting formulas describe the shifts of the osculating Keplerian orbital elements. Instead, the general relativistic effects are to be intended as written in terms of the nonosculating, contact elements, to the  first post--Newtonian order, because their disturbing functions depend on the velocity of the test particle. The corrections required to have final expressions in terms of the osculating elements are explicitly calculated. It turns out that, for a generally asymmetric range of values of the true anomaly smaller than the maximum allowed one, the gravitoelectric variations of the contact eccentricity, argument of pericentre and mean anomaly at epoch are different from their osculating counterparts. Instead, the gravitomagnetic orbital shifts,  written in terms of either the contact or the osculating elements, coincide if they are calculated over the entire unbound trajectory.}

The resulting formulas \textcolor{black}{for the Newtonian and post--Newtonian shifts, all expressed with the osculating elements and} valid for any spatial orientations of both the orbital plane and the spin axis of the source, are applied to ʻOumuamua, the first asteroid of interstellar origin which recently entered the inner regions of our solar system. While the quadrupole mass moment of the Sun and its angular momentum induce  shifts as little as a few microarcseconds throughout the whole path, the post--Newtonian gravitoelectric effect due to the solar mass amounts to less than a \textcolor{black}{hundred} milliarcseconds around the passage at perihelion. Actual usage of such kind of natural bodies as possible probes to perform gravitational experiments is likely prevented from the usually large non--conservative accelerations heavily perturbing their motions. As far as the Earth's flyby by the NEAR spacecraft is concerned, the size of its post--Newtonian disturbances is close to those of ʻOumuamua within \textcolor{black}{about} one order of magnitude. Instead, the perturbations due to the terrestrial oblateness are nominally much larger. However, the present--day relative uncertainty in our knowledge of the first even zonal harmonic of the geopotential is small enough to make the mismodeled part of such classical orbital shifts smaller than the corresponding relativistic ones.

The situation may become more favorable in the case of numerous planetary or satellite flybys by the many artificial probes currently travelling through the solar system. Furthermore, dedicated gravity experiments may be suitably designed relying upon the analytical results obtained.


The calculational approach adopted can straightforwardly be extended to any modified model of gravity as well.

\section*{Data availability}
No new data were generated or analysed in support of this research.
\section*{Conflict of interest statement}
I declare no conflicts of interest.
\bibliography{Megabib}{}
\end{document}